\documentclass[12pt]{iopart}

\usepackage{citesort}
\usepackage{graphicx}  % needed for figures
\usepackage{dcolumn}   % needed for some tables
\usepackage{bm}        % for math

\expandafter\let\csname equation*\endcsname\relax
\expandafter\let\csname endequation*\endcsname\relax
\usepackage{amssymb,amsmath}   % for math

\newcommand{\pd}[2]{\frac{\partial #1}{\partial #2}} 
\newcommand{\avg}[1]{\langle #1 \rangle}
\newcommand{\W}{\mathcal{W}}
\newcommand{\Qh}{\mathcal{Q}_{\rm h}}

\begin{document}
\rapid{Efficiency and Large Deviations in Time-Asymmetric Stochastic Heat Engines}
\author{Todd R. Gingrich$^1$, Grant M. Rotskoff$^{\hspace{2pt}2}$, Suriyanarayanan Vaikuntanathan$^{3}$, and Phillip L. Geissler$^{1, 4}$}
\address{$^1$Department of Chemistry, University of California, Berkeley, California 94720, USA}
\address{$^2$Biophysics Graduate Group, University of California, Berkeley, California 94720, USA}
\address{$^3$James Franck Institute and Department of Chemistry, University of Chicago, Chicago, Illinois 60637, USA}
\address{$^4$Materials Sciences Division and Chemical Sciences Division, Lawrence Berkeley National Laboratory, Berkeley, California 94720, USA}
\eads{\mailto{gingrich@berkeley.edu}, \mailto{rotskoff@berkeley.edu}, \mailto{svaikunt@uchicago.edu}, and \mailto{geissler@berkeley.edu}}

\begin{abstract}
In a stochastic heat engine driven by a cyclic non-equilibrium protocol, fluctuations in work and heat give rise to a fluctuating efficiency.  
Using computer simulations and tools from large deviation theory, we have examined these fluctuations in detail for a model two-state engine. 
We find in general that the form of efficiency probability distributions is similar to those described by Verley {\it et al.}~[2014 {\it Nat Comm}, {\bf 5} 4721], in particular featuring a local minimum in the long-time limit. 
In contrast to the time-symmetric engine protocols studied previously, however, this minimum need not occur at the value characteristic of a reversible Carnot engine. 
Furthermore, while the local minimum may reside at the global minimum of a large deviation rate function, it does not generally correspond to the least likely efficiency measured over finite time.
We introduce a general approximation for the finite-time efficiency distribution, $P(\eta)$, based on large deviation statistics of work and heat, that remains very accurate even when $P(\eta)$ deviates significantly from its large deviation form.
\end{abstract}

%\keywords{Stochastic Thermodynamics; Non-equilibrium Statistical Mechanics, Large Deviation Theory, Stochastic Engines}
\submitto{\NJP}

\section{Introduction}
\label{sec:intro}

As engineering capabilities reach molecular scales, design principles must account for the large fluctuations inherent in the behavior of nanoscale machines~\cite{Bennett1982,Lan2012,Li2007,Schindler2013,Browne2006,Leigh2003,Parrondo2002}.
These machines, ubiquitous in biology~\cite{Kinosita2000,Lau2007,Bustamante2001,Gelles1998} and increasingly relevant synthetically~\cite{Blickle2012,Hernandez2004}, operate stochastically.
As a consequence, familiar thermodynamic quantities, such as the heat absorbed from a hot bath $(\Qh)$, work extracted from the system $(-\W)$, and efficiency ($\eta \equiv -\W / \Qh)$, do not realize a single value.
Instead, each quantity varies from one measurement to the next according to a probability distribution and must be understood using stochastic thermodynamics~\cite{Esposito2010,Broeck2010,Crooks1999,Jarzynski2011,Verley2014}.
Fluctuations away from the mean behavior become insignificant in the long-time limit, but many molecular machines operate intermittently, performing their function over a short time.
In order to analyze the thermodynamic efficiency of nanoscale machines operating over a finite time, an understanding of the entire efficiency distribution is crucial.
Here, we aim to explore generic features of these efficiency distributions in non-equilibrium engines.

A system can be out of equilibrium in a time-independent manner if it is held in contact with multiple reservoirs maintained at different thermodynamic conditions (e.g.~\ two unequal temperature baths can induce a temperature gradient across a system). 
Verley {\it et al.}~\ constructed a model of one such system, a photoelectric cell, for which the temperature of Earth serves as one bath and the temperature of the Sun as another~\cite{Verley2014}. 
The non-equilibrium protocol driving these systems is time independent and therefore time-reversal symmetric. 
Verley {\it et al.}~\ employed large deviation theory to argue that in a long-time limit the efficiency distribution would attain the form $P(\eta) \sim e^{t_{\rm obs}J(\eta)}$, where $t_{\rm obs}$ denotes the observation time under these conditions.
Surprisingly, the large deviation rate function, $J(\eta)$, has a global minimum at the ``reversible'' efficiency, at which sufficiently long trajectories produce no entropy~\cite{Verley2014}.
In the case of a heat engine, the reversible efficiency is the Carnot efficiency, $\eta_{\rm C} = 1 - T_{\rm c}/T_{\rm h}$, the maximum that can be attained by a heat engine on average~\cite{Callen1960,Carnot1824}.

Systems can also be maintained out of equilibrium if they are driven in a time-dependent manner.
Standard engine protocols, including the Carnot and Stirling cycles, feature such time-dependent driving and generally lack time-reversal symmetry in both macroscopic and microscopic realizations~\cite{Callen1960,Blickle2012}.  
It is therefore important to understand efficiency distributions under much more generic cyclic driving protocols than those considered in reference~\cite{Verley2014}. 
Here we analyze engine performance fluctuations for this general case, exemplified by a model two-level heat engine.
Study of a similar model appeared shortly after the submission of our manuscript~\cite{Verley2014universal}.

Specifically, we examine the statistics and large deviation scaling for the work, heat, and thermodynamic efficiency for the dynamics sketched in figure~\ref{fig:schematic}. 
By computing a large deviation rate function for joint observations of $\W$ and $\Qh$, we in turn calculate the long-time behavior of the probability distribution for $\eta$~\cite{Touchette2009}.  
In particular, we determine the rate function for $\eta$, which resembles rate functions analyzed in~\cite{Verley2014}. 
However, we show that time-asymmetric driving shifts the location of the minimum away from the Carnot efficiency.
As such, the primary result of Verley {\it et al.}~\cite{Verley2014} does not generalize to encompass common engine protocols.

Furthermore, we note an important caveat pertaining to the relationship between the probability distribution of $\eta$ and its asymptotic representation as $e^{t_{\rm obs}J(\eta)}.$  As a consequence the efficiency distribution sampled over long but finite times may not reveal the minimum predicted by an analysis of the efficiency rate function.  
We obtain a general form of these finite-time distributions, which we expect to be significantly more relevant to the understanding of efficiency statistics in experiments.

\section{A Two-State Model Engine}
\begin{figure}[t!]
\begin{center}
\includegraphics[width=0.67\linewidth]{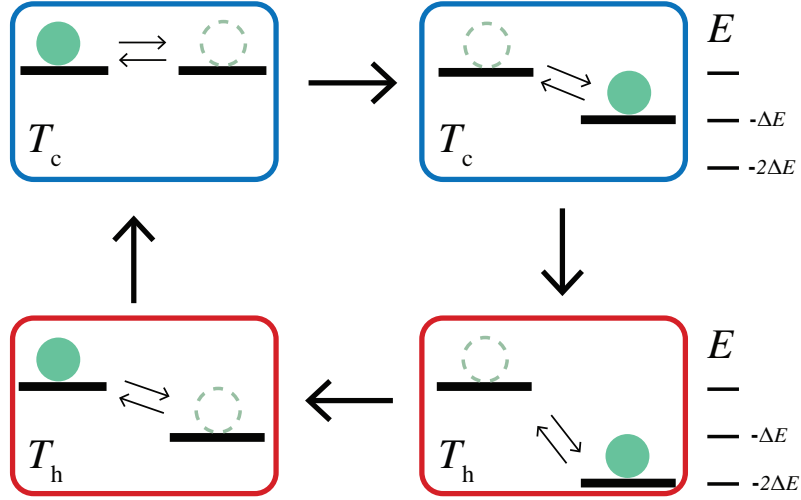}
\end{center}
\caption{A schematic of the model stochastic heat engine. 
A single particle can occupy one of two energy levels with thermal transitions between the states.  
The energy of the right state, $E_{\rm R}$, and the temperature, $T_{\rm h}$ or $T_{\rm c}$, are instantaneously switched in a cycle among four stages.
These switches occur in multiples of $\tau / 4$, where $\tau$ is the period of the cycle.}
\label{fig:schematic}
\end{figure}

We begin by constructing a two-state model of a stochastic engine. 
The temperature and energy levels are varied cyclically through four consecutive stages (see figure~\ref{fig:schematic}),
\begin{enumerate}
\item[]{Stage 1: \ $T^{(1)} = T_{\rm c}$, \ $E_{\rm L}^{(1)} = 0$, \ $E_{\rm R}^{(1)} = 0$ }
\item[]{Stage 2: \ $T^{(2)} = T_{\rm c}$, \ $E_{\rm L}^{(2)} = 0$, \ $E_{\rm R}^{(2)} = -\Delta E$ }
\item[]{Stage 3: \ $T^{(3)} = T_{\rm h}$, \ $E_{\rm L}^{(3)} = 0$, \ $E_{\rm R}^{(3)} = -2 \Delta E$ }
\item[]{Stage 4: \ $T^{(4)} = T_{\rm h}$, \ $E_{\rm L}^{(4)} = 0$, \ $E_{\rm R}^{(4)} = -\Delta E$, }
\end{enumerate}
with $E_{\rm L}$ and $E_{\rm R}$ the energies of the left and right states. 
$T_{\rm h}$ and $T_{\rm c}$ are the high and low temperatures respectively achieved by alternately coupling the two-state system to hot or cold baths. 
The superscript on temperatures and energies acts as an index for the stage of the cycle.

We carry out each cycle in time $\tau$, with each stage lasting for $\tau / 4$ units of time. 
During each stage the particle can hop between the left and right states with Arrhenius rates given by a tunable barrier height, $B$. 
The continuous time rate matrix for the $i^{\rm th}$ stage of the cycle is therefore 
\begin{equation}
\mathbb{W}^{(i)} = 
\begin{pmatrix}
-e^{-\beta^{(i)} B} & e^{-\beta^{(i)} (B - E_{\rm R}^{(i)})} \\ 
e^{-\beta^{(i)} B} & -e^{-\beta^{(i)} (B - E_{\rm R}^{(i)})}
\end{pmatrix},
\label{eq:ratematrix}
\end{equation}
with $\beta^{(i)} \equiv (k_{\rm B} T^{(i)})^{-1}$ and the Boltzmann constant, $k_{\rm B}$, set equal to unity throughout.

Work is extracted from the system when the right energy level is occupied while being instantaneously lowered. 
Each transition between the energy levels requires heat absorbed from the reservoir equal to the energy difference between the levels. 
We adopt the conventions that positive heat flow corresponds to heat flowing into the system and that positive work is performed on the system~\cite{Crooks1999,Jarzynski1997}. 
The simplicity of our four stage, two-state model lends itself to formal analysis of these fluctuating quantities as well as exhaustive computational study.

\section{Simulations}
\label{sec:sampling}

% sampling and LDP figure
\begin{figure}[t!]
\begin{center}
\includegraphics[width=0.95\linewidth]{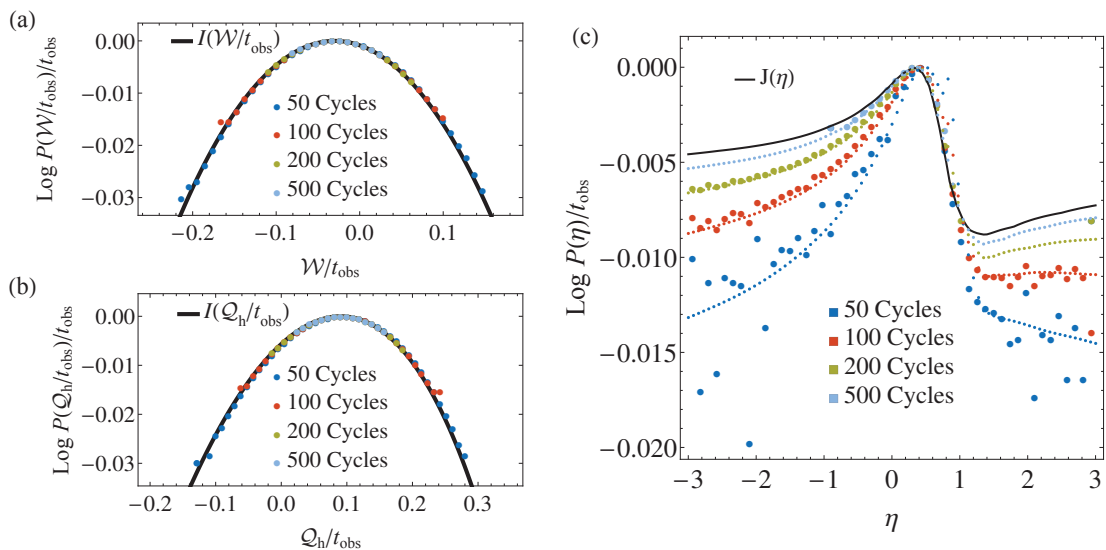}
\end{center}
\caption{(a) Work, (b) heat, and (c) efficiency sampling from $3 \times 10^7$ trajectories of the four stage, two-state engine with $\Delta E = 2.375, T_{\rm c} = 2, T_{\rm h} = 14, B = 0.05$, and period $\tau = 10$. 
The large deviation rate functions are shown as a solid black line. 
Histograms of sampled values are shown as large circles.  
The small circles in (c) plot the result of integrating out $\Delta S$ in equation~\eqref{eq:withjacobian}.}
\label{fig:sampling}
\end{figure}

While the particle dynamics of our model occurs in continuous time, the heat and work within each stage of a cycle are solely functions of the system's states at the beginning and end of that stage.
We advance time in units of $\tau / 4$ by drawing the state at the end of each stage in proportion to its exact probability, computed using the matrix exponential $\mathbb{T}^{(i)} = e^{\tau \mathbb{W}^{(i)}/4}.$ 
We collected statistics on the work extracted, heat absorbed, and efficiency of steady state stochastic trajectories evolved over many repeated engine cycles, focusing on a set of parameters ($\Delta E = 2.375, T_{\rm c} = 2, T_{\rm h} = 14, B = 0.05$, and $\tau = 10$) for which work is extracted on average.

Direct sampling of trajectories numerically illustrates that the heat and work distributions tend to a large deviation form as $t_{\rm obs} \to \infty$~\cite{Lebowitz1999,Touchette2009,Touchette2013}. 
The probability of observing a total work $\W$ and total heat absorbed from the hot bath $\Qh$ in a trajectory of length $t_{\rm obs}$ can thus be written as
\begin{equation}
\label{eq:ldt}
P(\Qh, \W) \sim e^{t_{\rm obs} I(\Qh /t_{\rm obs}, \W/t_{\rm obs})},
\end{equation}
where $\sim$ denotes an equality in the asymptotic limit and $I(\Qh/t_{\rm obs}, \W/t_{\rm obs})$ is the large deviation rate function. 
The large deviation scaling is robust even for a modest $t_{\rm obs}$ of only 50 engine cycles. 
The heat and work statistics of trajectories evolved for at least 50 engine cycles can therefore be well-described by a large deviation rate function.

The statistics of efficiency are also shown in figure~\ref{fig:sampling}(c) along with the efficiency rate function, $J(\eta)$. 
The sampled distributions for the reported finite-time measurements show no minimum. 
As expected, the efficiency distribution tends toward $P(\eta) \sim e^{t_{\rm obs} J(\eta)}$ at long times, but in practice even observation over 500 cycles is not sufficient for the large deviation rate function to predict the efficiency distribution. 
In marked contrast to the statistics of the time-additive quantities $\W$ and $\Qh$, the large deviation form is not predictive of the sampled efficiency distributions for the reported values of $t_{\rm obs}$.
 
\section{Large deviation rate functions for $\Qh, \W,$ and $\eta$}
\label{sec:rf2d} 

The large deviation function, $I(\Qh/t_{\rm obs}, \W/t_{\rm obs})$, can be calculated using standard methods of large deviation theory~\cite{Lebowitz1999,Touchette2009}. 
We introduce two fields, $\lambda_\W$ and $\lambda_{\Qh}$ and construct a scaled cumulant generating function for $\W$ and $\Qh$,
\begin{equation}
\psi(\lambda_{\Qh}, \lambda_{\W}) 
  = \lim_{t_{\rm obs} \to \infty} \frac{1}{t_{\rm obs}} \log \avg{e^{-\lambda_{\W}\W-\lambda_{\Qh}\Qh}}.
\label{eq:cgf}
\end{equation}
Applying a saddle point approximation, which is exact in the long-time limit, reveals that $I(\Qh/t_{\rm obs},\W/t_{\rm obs})$ can be obtained from $\psi(\lambda_{\Qh}, \lambda_{\W})$ by a Legendre transform.~\footnote{This logic assumes an absence of dynamic phase transitions. 
Otherwise the Legendre transform returns a convex hull of the rate function.}
In the long-time limit the scaled cumulant generating function can be found as a maximum eigenvalue of the appropriate ``tilted operator,'' which for our model must involve a product of tilted operators stemming from each stage of the engine~\cite{Touchette2009}.

\begin{figure}[t!]
\includegraphics[width=\linewidth]{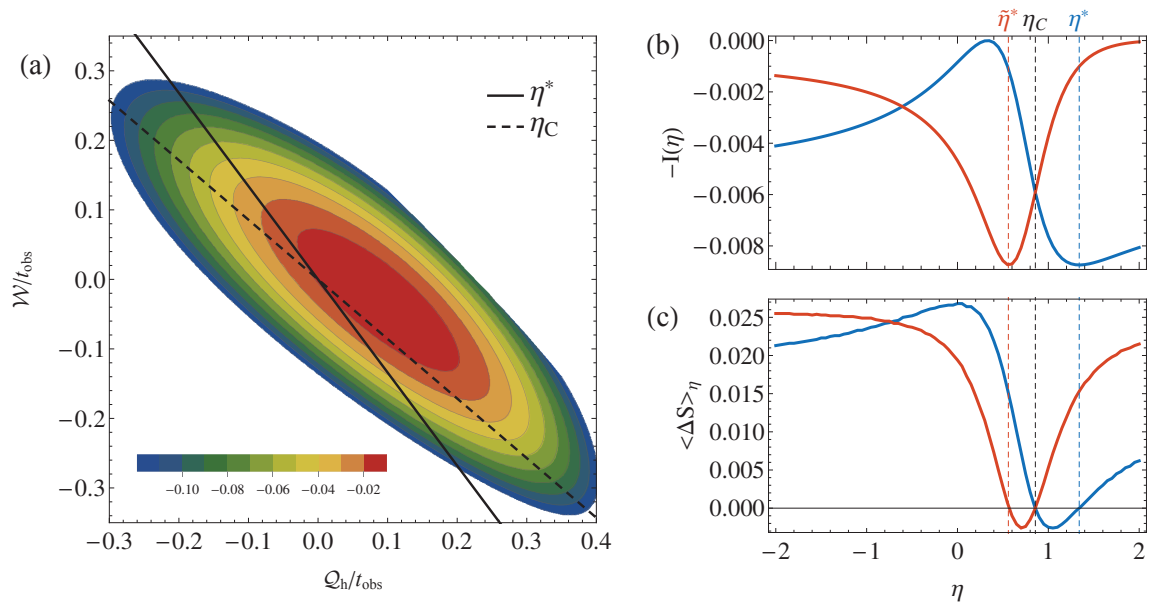}
\caption{(a) The large deviation rate function, $I$, for the joint observations of $\W$ and $\Qh$. 
Parameters for the engine protocol are the same as in figure~\ref{fig:sampling}. 
The black lines have slopes $-\eta^*$ and $-\eta_{\rm C}$, corresponding to the set of $(\Qh, \W)$ observations consistent with the respective values of efficiency. 
(b) The efficiency rate function evaluated at $\eta$ is given by the maximum of $I$ attained along a line through the origin with slope -$\eta$.  
Red and blue lines show the rate function for the forward and reverse protocols, respectively.  
Note that the minimal efficiency, $\eta^*$ corresponds to the slope of the tangent line through the origin in (a).
(c) Plot of the most likely value of entropy production in the long-time limit conditioned upon a given value of efficiency.  
Only in time-reversal-symmetric protocols will this function be strictly non-negative.}
\label{fig:rf2d}
\end{figure}

Recall that work is performed instantaneously between stages and only if the particle is in the right state, so we define the tilted operator
\begin{equation}
\label{eq:tiltedWork}
\mathbb{T}_{\W}^{(i)}(\lambda_{\W}) = 
\begin{pmatrix}
1 & 0 \\
0 & e^{-\lambda_{\W} \Delta E^{(i)}}
\end{pmatrix},
\end{equation}
whose derivatives with respect to $\lambda_{\W}$ provide statistical information about the work. 
$\Delta E^{(i)}$ denotes the change in $E_{\rm R}$ between stages of the protocol. 
The heat absorbed from the hot bath differs from the entropy production only by a factor of $\beta_{\rm h}$, so the tilted rate matrix for heat absorbed during those stages is analogous to the entropy production tilted operator of Lebowitz and Spohn~\cite{Lebowitz1999}. 
Its matrix elements are given by
\begin{equation}
\mathbb{W}_{\Qh}^{(i)}(\lambda_{\Qh})_{jk} = 
\begin{cases}
\mathbb{W}^{(i)}_{jk}, & j = k\\
(\mathbb{W}^{(i)}_{kj})^{T_{\rm h}\lambda_{\Qh}}(\mathbb{W}^{(i)}_{jk})^{1-T_{\rm h}\lambda_{\Qh}}, & j \neq k.
\end{cases}
\label{eq:tiltedQh}
\end{equation}
The tilted rate matrix propagates in continuous time, but it is convenient to also define a tilted operator which accounts for the complete stage, a time of $\tau / 4$:
\begin{equation}
\mathbb{T}_{\Qh}^{(i)}(\lambda_{\Qh}) = \exp\left(\frac{\tau}{4} \mathbb{W}^{(i)}_{\Qh}(\lambda_{\Qh})\right).
\label{eq:tiltedQhT}
\end{equation}
We now construct a tilted operator, $\mathbb{T}^{\rm cycle}$, for the entire cycle by forming a matrix product of the tilted operators for each stage (and each transition between stages).
\begin{equation}
\mathbb{T}^{\rm cycle}(\lambda_{\Qh}, \lambda_{\W}) = \mathbb{T}_\W^{(4)}(\lambda_\W) \mathbb{T}_{\Qh}^{(4)}(\lambda_{\Qh}) \mathbb{T}_\W^{(3)}(\lambda_\W) \mathbb{T}_{\Qh}^{(3)}(\lambda_{\Qh}) \mathbb{T}_\W^{(2)}(\lambda_\W) \mathbb{T}^{(2)} \mathbb{T}_\W^{(1)}(\lambda_\W) \mathbb{T}^{(1)}
\label{eq:cycletilted}
\end{equation}
Because we do not record heat absorbed from the cold bath, the cold stages involve time propagators rather than tilted operators for $\Qh$. 
$\mathbb{T}^{\rm cycle}$ raised to the $N^{\rm th}$ power generates the statistics of work and heat after $N$ cycles.
In the limit of a large number of cycles this matrix operator is dominated by the largest eigenvalue of $\mathbb{T}^{\rm cycle}$, which we denote as $\nu(\lambda_{\Qh}, \lambda_{\W})$. 
The scaled cumulant generating function can then be expressed as
\begin{equation}
\psi(\lambda_{\Qh},\lambda_{\W})  = \frac{1}{\tau} \log \nu(\lambda_{\Qh}, \lambda_{\W}).
\label{eq:psinu}
\end{equation}
A numerical Legendre transform gives the desired large deviation rate function, which is shown in figure~\ref{fig:rf2d}(a).

% 2d rate function figure

These long-time joint statistics of $\W$ and $\Qh$ determine the statistics of efficiency. 
We employ the contraction principle to obtain the efficiency rate function~\cite{Touchette2009}. 
This technique amounts to making a saddle-point approximation along each line of constant efficiency on $I,$
\begin{equation}
\label{eq:rfeta}
J(\eta) = \max_{\Qh} I(\Qh/t_{\rm obs}, -\eta \Qh/t_{\rm obs}).
\end{equation}
The result of the contraction is shown in figure~\ref{fig:rf2d}(b). 
As in the prior work of Verley et al.~\cite{Verley2014}, the minimum value obtained by each curve corresponds to a class of trajectories for which the average entropy production is zero, demonstrated in figure~\ref{fig:rf2d}(c). 
Our results demonstrate, however, that there can be two values of efficiency for which the average entropy production vanishes.
One of these values necessarily corresponds to $\eta_{\rm C}$.
For our engine it is the other efficiency value, $\eta^*$, for which $J(\eta)$ is minimal.

\section{The location of the minimum of $J(\eta)$}
\label{sec:theory}

The key distinguishing feature of our engine compared with those discussed in Ref.~\cite{Verley2014} is that our protocol lacks time-reversal symmetry. 
To emphasize the importance of time-reversal symmetry, we also plot the efficiency distribution of a time-reversed protocol, shown in red in figure~\ref{fig:rf2d}. 
The minimum in the rate function for the time-reversed protocol, denoted $\tilde{\eta}^*$, is distinct from both $\eta^*$ and $\eta_{\rm C}$. 
Indeed, we can relate $\eta^*$ to $\tilde{\eta}^*$ via the fluctuation theorem~\cite{Crooks1999,Seifert2005}, thereby illustrating that $\eta^* = \eta_{\rm C}$ if and only if the protocol is time-reversal symmetric.

Entropy production is defined on a single trajectory level as
\begin{equation}
\Delta S = \log \frac{P[x(t)|\Lambda(t)]}{\tilde{P}[\tilde{x}(t)|\tilde{\Lambda}(t)]},
\label{eq:ft}
\end{equation}
where $x(t)$ is a trajectory subject to a time-dependent non-equilibrium protocol $\Lambda(t)$ and a tilde denotes the time reversal of a function~\cite{Seifert2012}. 
Assuming that the dynamics of the system is microscopically reversible, we can interpret the entropy production in terms of the heat absorbed from the thermal reservoirs. 
In particular, 
\begin{equation}
\Delta S = -\beta_{\rm h} \Qh - \beta_{\rm c} \mathcal{Q}_{\rm c} + \log
\frac{p_{\rm ss}(x(0)|\Lambda(0))}
     {\tilde{p}_{\rm ss}(\tilde{x}(0)|\tilde{\Lambda}(0))},
\label{eq:ds_heat}
\end{equation}
with $p_{\rm ss}$ denoting the steady-state probability~\cite{Crooks1999}. 
The heat is extensive in time; the contribution involving $p_{\rm ss}$ and its time-reversed counterpart is subextensive for the two-state system and can therefore be neglected in the long-time limit.
By the first law, $\mathcal{Q}_{\rm c} = \Delta E - W - \Qh$.
Here we can neglect $\Delta E$ which also grows sub-extensively with the length of the trajectory.
Thus in the long-time limit $\Delta S = \beta_{\rm c} \Qh \left(\eta_{\rm C} - \eta\right)$.
It follows from equation~\ref{eq:ft} that
\begin{equation}
I(\Qh/t_{\rm obs},\W/t_{\rm obs}) = \tilde{I}(-\Qh/t_{\rm obs},-\W/t_{\rm obs}) + \beta_{\rm c} \left(\Qh \eta_{\rm C} + \W\right)/t_{\rm obs},
\label{eq:ftI}
\end{equation}
where $\tilde{I}$ is the large deviation rate function for the time-reversed protocol.

Geometrically, the set of work and heat values yielding efficiency $\eta$ fall on a line in the $(\Qh/t_{\rm obs}, \W/t_{\rm obs})$ plane passing through the origin with slope $-\eta$.
Lines corresponding to efficiencies $\eta^*$ and $\eta_{\rm C}$ are drawn in figure~\ref{fig:rf2d}.
In the long-time limit the probability of observing a given efficiency is dominated by the most likely point on this line.
Thus $J(\eta)$ is extracted from the maximum of $I(\Qh/t_{\rm obs}, \W/t_{\rm obs})$ along the line with slope $-\eta$, as expressed in equation~\eqref{eq:rfeta}.
The minima of the efficiency rate functions therefore correspond to the lines tangent to the level curve of $I$ at $(\Qh, \W) = (0, 0)$, requiring
\begin{equation}
\eta^* = \frac{\left.\left(\pd{I}{\Qh}\right)\right|_{(0,0)}}{\left.\left(\pd{I}{\W}\right)\right|_{(0,0)}} \hspace{0.1\linewidth} \tilde{\eta}^* = \frac{\left.\left(\pd{\tilde{I}}{\Qh}\right)\right|_{(0,0)}}{\left.\left(\pd{\tilde{I}}{\W}\right)\right|_{(0,0)}}.
\label{eq:etastars}
\end{equation}
Equation~\eqref{eq:etastars} implies $\left.\left(\pd{I}{\Delta S}\right)\right|_{\stackrel{\Delta S = 0}{\eta = \eta^*}} = 0$ and $\left<\Delta S\right>_{\eta^*} = 0$, where the average is taken over the subensemble of trajectories with efficiency $\eta^*$.
Differentiation of equation~\eqref{eq:ftI} with respect to $\W$ and $\Qh$ implies, after some simplifying algebra,
\begin{equation}
\eta^*-\eta_{\rm C} = \left(\eta^* - \tilde{\eta}^*\right) t_{\rm obs} T_{\rm c}\left.\left(\pd{\tilde{I}}{\W}\right)\right|_{(0,0)}.
\label{eq:etadiff}
\end{equation}
Note that the derivative $\left.\pd{\tilde{I}}{\W}\right|_{(0, 0)}$ is non-zero because (i) $I$ and $\tilde{I}$ are convex, (ii) their maxima locate the corresponding mean values of $\W$ and $\Qh$, and (iii) a useful engine should extract nonzero work on average.
We therefore see that $\eta^* =\eta_{\rm C}$ precisely when $\eta^* = \tilde{\eta}^*$. 
The minimum of $J(\eta)$ occurs at the Carnot efficiency if the protocol is time-reversal symmetric since the symmetry enforces $\eta^* = \tilde{\eta}^*$. 
In the more generic case of time-asymmetric engines, however, distinct values of $\eta^*$ and $\tilde{\eta}^*$ imply that neither of the minima occur at $\eta_{\rm C}$.

\section{Finite time efficiency distributions}

As demonstrated by numerical simulation, there is a significant regime of observation times for which the efficiency distribution is not well-described by the large deviation form and no local minimum is evident. 
The minimum in the efficiency rate function therefore may not be apparent for actual experimental measurement of efficiency statistics. 
Nevertheless, we may leverage the large deviation form for work and heat, equation~\eqref{eq:ldt}, to construct an approximation for $P(\eta)$ that is much more faithful to finite-time statistics.

Consider the coordinate transformation from $(\Qh, \W)$ to $(\eta, \Delta S)$, where
\begin{equation}
\eta = -\frac{\W}{\Qh}, \hspace{0.1\linewidth} \Delta S = -\beta_{\rm c}(\W + \eta_{\rm C} \Qh).
\label{eq:transformationdef}
\end{equation}
The Jacobian for this transformation contributes negligibly to the distribution $P(\eta, \Delta S)$ in the long-time limit since it does not vary exponentially with $t_{\rm obs}$. 
At long but finite times, however, it can strongly shape statistics of $\eta$ and $\Delta S$. 
Retaining this Jacobian, while exploiting the large deviation form of $P(\Qh, \W)$, we estimate 
\begin{equation}
P(\eta, \Delta S) \sim \frac{ T_{\rm c}^2 \left|\Delta S\right|}{(\eta - \eta_{\rm C})^2} e^{t_{\rm obs} I\left(\frac{T_{\rm c} \Delta S }{(\eta - \eta_{\rm C})t_{\rm obs}},\frac{\eta T_{\rm c} \Delta S}{(\eta_{\rm C} - \eta)t_{\rm obs}}\right)}.
\label{eq:withjacobian}
\end{equation}
Equation~\eqref{eq:withjacobian} is a very general result for the joint distribution of efficiency and entropy production.
Its sole underlying assumption is that work and heat fluctuations are well-described by a large deviation form.
For a model dynamics in which $\W$ and $\Qh$ obey a large deviation principle at all times, such as that studied in~\cite{Polettini2014}, equation~\eqref{eq:withjacobian} is thus exact.

Obtaining an efficiency distribution $P(\eta)$ from this result requires marginalizing over $\Delta S$.
In work that appeared after the submission of this manuscript, Polettini et al.\ in effect integrated equation~\eqref{eq:withjacobian} over $\Delta S$ analytically for a linear response model whose work and heat statistics have a large deviation form by construction~\cite{Polettini2014}.
More generally, the assumption leading to equation~\eqref{eq:withjacobian} will be valid only when $t_{\rm obs}$ is sufficiently large.
In this case a saddle point approximation is likely to be a well-justified and practical alternative to exact marginalization.
For our two-state system this alternative approach produces very accurate predictions for $P(\eta)$ at finite $t_{\rm obs}$ (small dotted lines in figure~\ref{fig:sampling}(c)).
The form of $P(\eta)$ may therefore be reliably extracted from the large deviation form for work and heat fluctuations.

The prefactor in equation~\eqref{eq:withjacobian} attenuates the probability of observing very large positive and negative values of $\eta$ when $t_{\rm obs}$ is finite.
As a result, for any finite-time observation $\eta^*$ will not strictly be the \emph{least} likely; $\eta$ is a continuous variable with infinite support, so no finite efficiency can be the least probable.
Nevertheless, at very long times, trajectories with efficiency $\eta^*$ will be increasingly rare. 

\section{Discussion}
\label{sec:discussion}
% Macroscopic heat engines
Efficiency is meant to provide an assessment of how much work can be extracted from a machine relative to the expense of operating it. 
For macroscopic heat engines, fluctuations in work and heat are vanishingly small compared to their means such that work, heat, and efficiency can be reasonably replaced by their average values.
In contrast, fluctuations cannot be neglected for a microscopic engine.
An understanding of the finite-time and long-time statistics of the efficiency provides a lens through which to assess the design of a microscopic engine.

We have extended the analysis of efficiency fluctuations~\cite{Verley2014} to include the common case of time-asymmetric driving, illustrating that the Carnot efficiency does not minimize the efficiency rate function for most molecular machines.
Under both time-symmetric and time-asymmetric driving, long trajectories that realize the Carnot efficiency necessarily have zero entropy production by equation~\eqref{eq:transformationdef}.
However, in the case of time-asymmetric driving, the minimum in the efficiency rate function corresponds to a value $\eta^*$, distinct from $\eta_{\rm C}$.
The subensemble of trajectories with efficiency $\eta^*$ has a vanishing entropy production \emph{on average} as shown in figure~\ref{fig:rf2d}.
For both $\eta^*$ and $\eta_{\rm C}$, $\left<\Delta S\right>_\eta = 0$, complicating the notion of a reversible efficiency.

For sufficiently long observation times, a local minimum is evident in the efficiency distribution.
A minimum in the efficiency large deviation rate function may provide theoretical insight, but will be irrelevant to many experiments.
We have demonstrated that finite-time efficiency distributions can be accurately captured using a large deviation form for joint fluctuations of heat and work.
We anticipate that the form of these distributions will be a more relevant consideration for the design of microscopic engines than the identification of $\eta^*$.

\section*{Acknowledgments}
We acknowledge support from the Fannie and John Hertz Foundation (T.R.G.) and the National Science Foundation Graduate Research Fellowship (G.M.R.). 
This work was supported in part by the Director, Office of Science, Office of Basic Energy Sciences, Materials Sciences, and Engineering Division, of the U.S. Department of Energy under contract No.\ DE AC02-05CH11231 (S.V. and P.L.G.). 
S.V. also acknowledges support from The University of Chicago.

\section*{References}
%cinclude the bbl file for arxiv submission
\providecommand{\newblock}{}

\end{document}